\documentclass{aa}
\usepackage{txfonts,natbib,graphicx}
\bibpunct{(}{)}{;}{a}{}{,}

\newcommand{\irnulvijf}{IRAS\,05341+0852}
\newcommand{\irnulzes}{IRAS\,06530$-$0213}
\newcommand{\irnulacht}{IRAS\,08143$-$4406}

\renewcommand{\aap}{A\&A}
\renewcommand{\aaps}{A\&AS}
\renewcommand{\apj}{ApJ}
\renewcommand{\apjl}{ApJL}
\renewcommand{\araa}{ARA\&A}
\renewcommand{\mnras}{MNRAS}
\renewcommand{\nat}{Nat.}
\renewcommand{\solphys}{Sol. Phys.}
\renewcommand{\pra}{Phys. Rev. A}

\begin{document}
\title{Detection of elements beyond the Ba-peak\\
in VLT+UVES spectra of post-AGB stars\thanks{Based on
observations collected at the European Southern Observatory, Paranal, Chile
(ESO Programme 66.D-0171)}}

\author{Maarten Reyniers
\and Hans Van Winckel\thanks{Postdoctoral fellow of the Fund for Scientific Research, Flanders}}
\institute{Departement Natuurkunde en Sterrenkunde, K.U.Leuven, Celestijnenlaan 200B, 3001 Leuven, Belgium}
\offprints{Maarten Reyniers, \email{Maarten.Reyniers@ster.kuleuven.ac.be}}
\date{Received 14 May 2003 / Accepted 1 August 2003}

\abstract{In this letter, we report on our successful systematic search for
lines of elements beyond the Ba-peak in spectra of s-process enriched post-AGB
stars. Using newly released atomic data from both the VALD database and the
D.R.E.A.M. project, we could derive abundances for several elements heavier
than europium for three objects, on the base of high quality VLT+UVES spectra.
The abundances of these elements are of particular interest since they
turn out to be powerful constraints for chemical evolutionary AGB models.
Their high abundances indicate that, also in only moderately metal deficient
AGB stars, production of lead is expected to be significant.
}

\authorrunning{M. Reyniers \& H. Van Winckel}
\titlerunning{Elements beyond the Ba-peak in post-AGB stars}

\maketitle

\section{Introduction}
There is now general consensus that the neutron nucleosynthesis in AGB stars
is triggered by proton engulfment in the $^{12}$C rich intershell
\citep[e.g.][]{busso99}. A significant $^{13}$C-pocket is formed in a limited
p/$^{12}$C regime and the neutron irradiation occurs in radiative conditions
\citep{straniero95} during the interpulse period, when this $^{13}$C pocket
is consumed by $\alpha$ captures producing $^{16}$O and free neutrons. The
dredge-ups associated with the thermal pulses during the AGB evolution will
bring the enriched material to the stellar surface. There is, however, no
consensus on the mechanisms of both the dredge-up processes and the partial
mixing mechanism of protons in the intershell
\citep[e.g.][ and references therein]{lugaro03}. 

The characteristics of the s-process nucleosynthesis are expected to be
strongly dependent on the metallicity since the $^{12}$C in the intershell
is of primary origin, and thus largely independent on the initial $^{12}$C
content.  Moreover, since the p/$^{12}$C regime for which  the s-process is
expected to be active, is limited \citep{goriely00}, the uncertainties on the 
neutron irradiation are mainly from the extend of the proton engulfed zone
in the intershell, and much less on the ad-hoc assumptions of the proton 
profile. In reduced environments, there are more neutrons available per iron
seed, pushing the s-process to heavier elements. The prediction that in metal
deficient ([Fe/H]\,$<$\,$-$1.0) stars, the production of Pb and Bi is expected
to be significant \citep{gallino98, goriely00} was recently confirmed by the
detection of such `lead-stars' in extrinsic CH-stars by \citet{vaneck01}. 

The simple picture, in which the metallicity is the main parameter determining
the s-process characteristics is, however, much less clear after the finding
of a number of objects of the same metallicity, but with a large variety of
s-process characteristics: s-process enriched objects of low metallicity
($-$3\,$<$\,[Fe/H]\,$<$\,$-$1) were found without the expected Pb enrichment 
\citep{aoki01, vaneck03}; and in less deficient post-AGB objects, large
overabundances were found in which again a large spread in s-process efficiency
was observed at similar metallicity \citep{vanwinckel00, reyniers03}.

In this paper we focus on a systematic search of the very heavy elements beyond
the Ba-peak for the intrinsic post-AGB stars. The reason for this search is
twofold: an accurate distribution of s-process overabundances is a prime tool
to constrain the s-process nucleosynthesis and, second, in two, somewhat hotter
post-AGB stars, hafnium (Z\,=\,72) lines were tentatively detected, yielding a
relatively large overabundance \citep{vanwinckel00}. Moreover, new atomic data
for s-process elements beyond the Ba-peak became recently available, either
through the VALD database \citep{kupka99}, or via the new D.R.E.A.M. project
(http://www.umh.ac.be/$\sim$astro/dream.shtml). We used our high-resolution,
high signal-to-noise (S/N) VLT+UVES spectra for this purpose. These spectra are
described in detail in \citet{reyniers03}. Here, we only note that they
cover a spectral interval from 3745\,\AA\ to 10550\,\AA; that their
resolving power varies between 55,000 and 60,000; and that the S/N is lying
between 100 and 200 (depending on wavelength).

\section{Systematic search}
\subsection{\irnulzes}
We first concentrated on the spectrum of the heavily enriched \irnulzes\ 
\citep{reyniers03}. We extracted all atomic data for lines of elements
between gadolinium (Gd, Z\,=\,64) and platinum (Pt, Z\,=\,78) from either
the VALD database or the D.R.E.A.M. project. By assuming an ad-hoc high
abundance, we determined a list of the strongest lines of all elements using
a model atmosphere of \irnulzes. The eventual presence of those lines was
checked on the spectrum until the estimated strengths became too small to
detect.

After that, the detected lines were thoroughly checked in four different ways:
{\em(1)} Lines with a suspected profile (asymmetry, continuum placement, cosmic
hit, end of CCD) were immediately discarded; {\em(2)} Lines were compared to the
spectrum of HR\,4656, a fast rotating
(V$\sin i$\,$\simeq$\,\,210\,km\,s$^{-1}$) Beta Cephei with spectral type
B2IV, in order to check for telluric blending. Also the solar spectrum
identification by \citet{moore66} was used for this purpose; {\em(3)} We also
checked for possible diffuse interstellar bands (DIBs) in proximity of the line
under study; {\em(4)} Blends are not always ``visible'' by a simple inspection
of the line profile. Therefore, a final check was performed on the lines by the
use of the VALD database. We extracted all lines from VALD in the immediate
vicinity ($\pm$1\,\AA) of the line under study, and calculated the equivalent
widths of all these extracted lines, using the abundances of \irnulzes. Such an
exercise is highly necessary since it reveals sometimes strong blends (mostly
from other s-process lines) which were not detected at first glance.

The final selection for \irnulzes\ contains three lines of singly ionised
gadolinium (Gd\,{\sc ii}, Z\,=\,64), two lines of singly ionised ytterbium
(Yb\,{\sc ii}, Z\,=\,70) and two surprisingly strong lines of singly ionised
lutetium (Lu\,{\sc ii}, Z\,=\,71). We also found one line of tungsten
(W, Z\,=\,74), but the line is blended with a Sm\,{\sc ii} line. Nevertheless,
we decided to derive a W abundance from this line by using spectrum
synthesis (see Sect. \ref{sect:elementsbnd}). These lines are all
shown in Fig.~\ref{fig:smoverlplijn}; their atomic data and equivalent widths
can be found in Tab.~\ref{tab:ndvdllns}.

\vspace{-.17cm} \subsection{\irnulvijf} \vspace{-.17cm}
All lines detected in \irnulzes\ were obviously also detected in the
VLT+UVES spectrum of \irnulvijf, but, as can be seen on
Fig.~\ref{fig:smoverlplijn} and Tab.~\ref{tab:ndvdllns}, not all lines
were used for an abundance determination since some are too blended with
other s-process lines (e.g. the Gd\,{\sc ii} line at 5092.249\,\AA) or some
are even too strong (e.g. the blended W\,{\sc ii} line at 5104.432\,\AA). On
the other hand, three additional Gd\,{\sc ii} lines and one additional
Yb\,{\sc ii} line were found suitable for an abundance determination. Plots
and atomic data can again be found in Fig.~\ref{tab:ndvdllns} and
Tab.~\ref{tab:ndvdllns} respectively.

\vspace{-.17cm} \subsection{\irnulacht} \vspace{-.17cm}
The line strengths of the trans-Ba elements detected in \irnulzes\ and
\irnulvijf\ are much smaller for \irnulacht, due to the less extreme
enhancement of this object \citep{reyniers03}. Some lines are even untraceable
(e.g. the Gd\,{\sc ii} line at 5469.713\,\AA). As a result, only three lines
were found suitable for an abundance determination.

\begin{table*}
\caption{Atomic data and abundances of individual lines of s-process elements
with Z\,$>$\,63.}\label{tab:ndvdllns}
\begin{center}
\begin{tabular}{cccccccccc}
\hline
 & & & & \multicolumn{2}{c}{IRAS\,05341} & \multicolumn{2}{c}{IRAS\,06530} & \multicolumn{2}{c}{IRAS\,08143}\\
$\lambda$ & $\chi$ & $\log gf$ & ref. & W$_{\lambda}$ & $\log \epsilon$ & 
 W$_{\lambda}$ & $\log \epsilon$ & W$_{\lambda}$ & $\log \epsilon$ \\
(\AA) & (eV) & & & (m\AA) & & (m\AA) & & (m\AA) \\
\hline
\multicolumn{10}{c}{Gd\,{\sc ii}}\\
\hline
 5092.249 & 1.727 & $-$0.529 & (1) &\ldots&\ldots& 45.6 & 2.72 &  9.5 & 2.04 \\
 5108.903 & 1.659 & $-$0.322 & (1) & 66.2 & 2.26 & 40.5 & 2.39 &  9.7 & 1.78 \\
 5357.794 & 1.157 & $-$1.470 & (1) & 36.5 & 2.59 &\ldots&\ldots&\ldots&\ldots\\
 5469.713 & 1.102 & $-$1.540 & (1) & 33.8 & 2.56 &\ldots&\ldots&\ldots&\ldots\\
 5733.852 & 1.372 & $-$0.893 & (1) & 63.6 & 2.52 & 36.4 & 2.64 &\ldots&\ldots\\
 6080.641 & 1.727 & $-$0.926 & (1) & 29.9 & 2.42 &\ldots&\ldots&\ldots&\ldots\\
\hline
\multicolumn{10}{c}{Yb\,{\sc ii}}\\
\hline
 5352.954 & 3.747 & $-$0.340 & (2) &\ldots&\ldots& 54.8 & 2.59 &\ldots&\ldots\\
 5651.988 & 3.747 & $-$0.800 & (2) & 62.9 & 2.88 & 38.3 & 2.84 &\ldots&\ldots\\
 6489.298 & 4.153 & $-$0.810 & (2) & 48.2 & 3.07 &\ldots&\ldots&\ldots&\ldots\\
\hline
\multicolumn{10}{c}{Lu\,{\sc ii}}\\
\hline
 6221.890 & 1.542 & $-$0.760 & (3) & 245. & 1.58$^\star$ & 187.4 & 1.74$^\star$ & 42.1 & 0.97\\ 
 6463.107 & 1.463 & $-$1.050 & (3) & 230. & 1.53$^\star$ & 157.9 & 1.78$^\star$ &\ldots&\ldots\\
\hline
\multicolumn{10}{c}{W\,{\sc ii}}\\
\hline
 5104.432 & 2.355 & $-$0.910 & (4) &\ldots&\ldots& 73.6 & 2.92$^\dag$ &\ldots&\ldots\\
\hline
\multicolumn{10}{l}{Refs. (1) \citealt{meggers75} via \citealt{kurucz93} and VALD
(2) \citealt{biemont98} via D.R.E.A.M.}\\
\multicolumn{10}{l}{(3) \citealt{quinet99} via D.R.E.A.M.
(4) \citealt{clawson73} via \citealt{kurucz93} and VALD}\\
\multicolumn{10}{l}{$^\star$abundance derived with spectrum synthesis (see text)
$^\dag$Sm blend taken into account (see text)}\\
\end{tabular}
\end{center}
\vskip -.5 cm
\end{table*}

\vspace{-.2cm}\section{Abundances}\label{sect:elementsbnd}
Tab.~\ref{tab:ndvdllns} also contains the abundances derived from the
selected lines. These abundances were calculated using Kurucz' model atmospheres
\citep{kurucz93} and the latest version (April 2002) of Sneden's Stellar Line
Analysis Program MOOG. Model parameters were taken from our earlier studies
(\irnulvijf: \citealt{vanwinckel00}; \irnulzes\ and \irnulacht:
\citealt{reyniers03}).

\begin{figure*}
\resizebox{\hsize}{!}{\rotatebox{-90}{\includegraphics{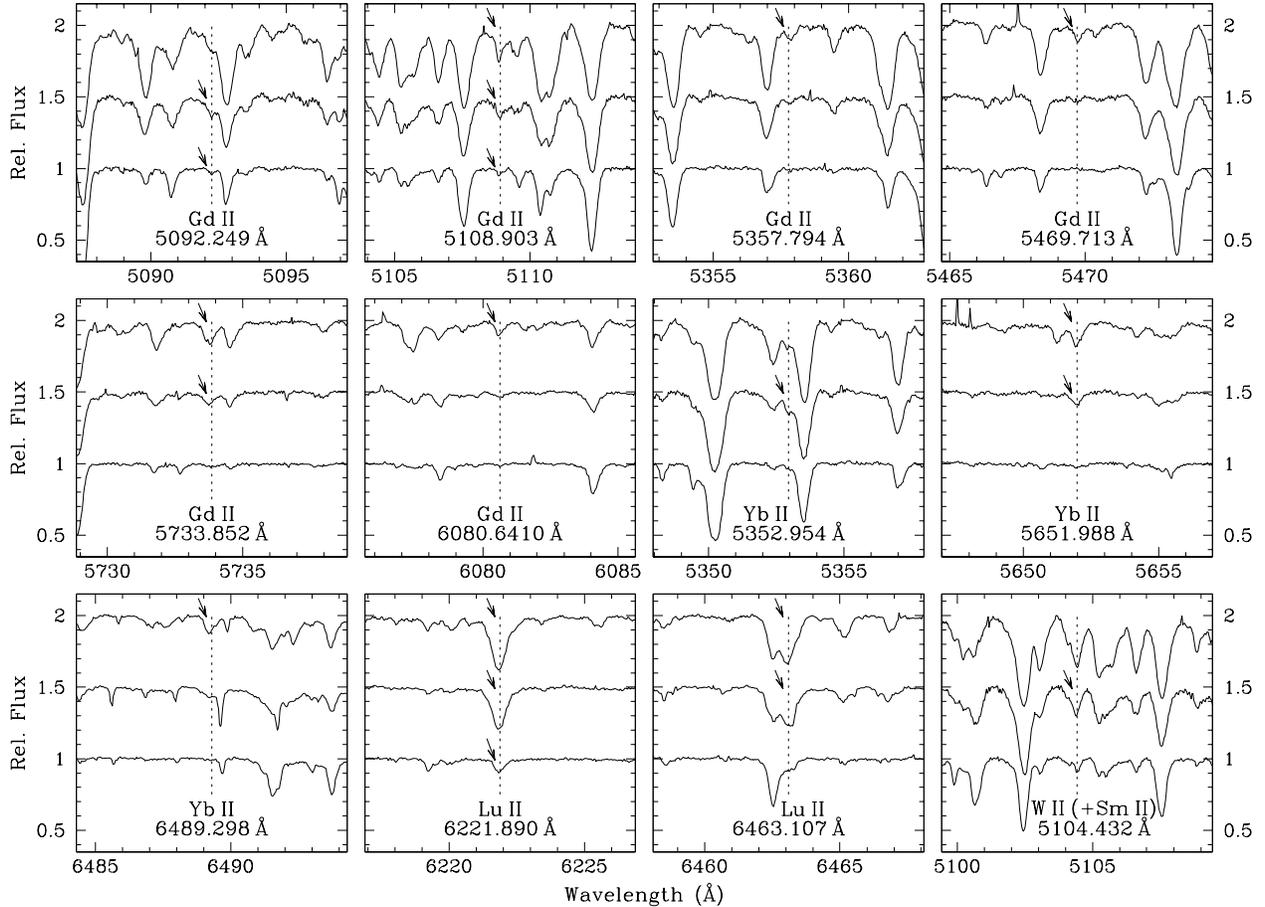}}}
\vskip -0.9 cm
\caption{Detection of lines from elements beyond the Ba-peak in the
spectra of \irnulvijf, \irnulzes\ and \irnulacht. All these lines are
intensively checked for blends, telluric pollution and DIB contamination
(see text for details). From top to bottom are plotted: \irnulvijf, \irnulzes\ 
and \irnulacht. The spectral lines marked with an arrow are used for abundance
determination.}\label{fig:smoverlplijn}
\end{figure*}

The W line at 5104.432\,\AA\ does not fulfill the criteria mentioned
above since it is blended with a Sm\,{\sc ii} line at 5104.479\,\AA.
In the case of \irnulzes, the total equivalent width of the blend is
73.6\,m\AA, and the predicted contribution of the Sm\,{\sc ii} line is
22\,m\AA\ (assuming $\log\epsilon$(Sm)\,=\,2.09). Spectrum synthesis with
a macroturbulent broadening $\xi_m$\,=\,19.5\,km\,s$^{-1}$
\citep[see][]{reyniers03} yields a W abundance of $\log \epsilon$(W)\,=\,3.04.
Since it concerns an isolated blend, we deduced the abundance also without
profile match solely on the equivalent width basis. This technique yields an
abundance of $\log \epsilon$(W)\,=\,2.92. We prefer to use this lower
abundance since it is independent of the adopted $\xi_m$. However, one
has to treat this abundance with caution since there might also be a blend
present in the blue wing of the profile. We could not identify this blend, but
it seems also present in the spectrum of the two other objects (see 
Fig.~\ref{fig:smoverlplijn}). In the case of \irnulvijf, we did not derive a
W abundance from the blended line at 5104.432\,\AA, due to the relatively high
contribution of the Sm\,{\sc ii} line in the blend (80.5\,m\AA\ assuming
$\log\epsilon$(Sm)\,=\,2.20). Also for \irnulacht, the relative contribution
is considered to be too high for a reliable abundance determination.

For Lu, hyperfine splitting (hfs) of the energy levels is expected. The Lu
abundance was therefore deduced by spectrum synthesis. For the line at
6221.890\,\AA, there are nine hfs components. Hfs A and B constants for the
lower level were taken from \citet{brix52}; for the upper level from
\citet{denhartog98}. For the 6463.107\,\AA\ transition, there are only three
components, and the splitting is entirely due to the lower level, since for
the upper level $J$\,$=$\,$0$. Hfs constants for this level were also taken
from \citet{denhartog98}. Relative strengths were calculated with the
equations given in \citet{condon35}. The effect of hfs on the Lu abundance
determination is considerable. In the case of \irnulzes, the difference in
abundance between a hfs and a non-hfs synthesis ($\Delta$hfs) is $-$0.27\,dex
for the 6221.890\,\AA\ line, and $-$0.14\,dex for the 6463.107\,\AA\ line.
The effect is even larger for the cooler \irnulvijf: $\Delta$hfs\,=\,$-1.37$
and $-$0.48 for the two lines respectively. For the hotter and less enhanced
\irnulacht, hfs only affects the line profile, but not the equivalent width.

In Tab.~\ref{tab:ultrhvy050608} a summary of the final abundances is given.
In order to compare these abundances with the solar ones, it should be
checked that the solar abundances are based on the same sources of $gf$
values as used in the present paper. This is, however, for several reasons
not always straightforward. The original reference for the solar Gd abundance
is \citet{bergstrom88} and it is based on eight Gd\,{\sc ii} lines, without
overlap with the lines in this paper. However, the $\log gf$ values of these
eight lines are in our source \citep{meggers75} somewhat lower (on average
0.14\,dex) than the \citeauthor{bergstrom88} values. This might be an
indication that the \citeauthor{meggers75} values are systematically too
low and that the Gd abundances should be lowered by $\sim$0.14\,dex to be
consistent with the solar value. For Yb, the details of the solar abundance
calculation are unfortunately lost (Grevesse, private communication). The
solar Lu abundance is derived from the 6221\,\AA\ line \citep{bord98,
denhartog98}, with the same $\log gf$ and hfs parameters as in this paper.
Finally, the situation is less clear for W, since the solar value is derived
from {\em neutral} W lines \citep{holweger82, duquette81}. Moreover, there is
a large discrepancy (0.42\,dex) between the photospheric and the meteoritic
value for this element, the photospheric being the most uncertain one
\citep{grevesse98}.

\begin{table}
\caption{Final abundances of elements beyond the Ba-peak.
For each ion, we list the number of lines N, the mean equivalent width
$\overline{W_{\lambda}}$, the absolute abundance $\log\epsilon$ (i.e. relative
to H: $\log\epsilon$\,=\,$\log{\rm X/H}$\,+\,12), the line-to-line scatter
$\sigma$ and the abundance relative to iron [el/Fe]. ss stands for spectrum
synthesis, bl means corrected for blend. (Photospheric) solar abundances are
taken from the review by \citet{grevesse98}:
$\log\epsilon$(Gd)$_{\odot}$\,=\,1.12,
$\log\epsilon$(Yb)$_{\odot}$\,=\,1.08,
$\log\epsilon$(Lu)$_{\odot}$\,=\,0.06,
$\log\epsilon$(W)$_{\odot}$\,=\,1.11.}\label{tab:ultrhvy050608}
\begin{center}
\begin{tabular}{lccccc}
\hline
ion & N &{\rule[0mm]{0mm}{4mm}$\overline{W_{\lambda}}$}&$\log\epsilon$&$\sigma$&[el/Fe]\\
\hline
\multicolumn{6}{c}{\rule[0mm]{0mm}{4mm}\irnulvijf}\\
\multicolumn{6}{c}{\rule[-2mm]{0mm}{4mm}[Fe/H]\,=\,$-$0.7, T$_{\rm eff}$\,=\,6500\,K, $\log g$\,=\,1.0}\\
\hline
Gd\,{\sc ii} & 5 & 46 & 2.47 & 0.13 &  2.07\\
Yb\,{\sc ii} & 2 & 56 & 2.98 & 0.13 &  2.62\\
Lu\,{\sc ii} & 2 & ss & 1.56 & 0.04 &  2.22\\
\hline      
\multicolumn{6}{c}{\rule[0mm]{0mm}{4mm}\irnulzes}\\
\multicolumn{6}{c}{\rule[-2mm]{0mm}{4mm}[Fe/H]\,=\,$-$0.5, T$_{\rm eff}$\,=\,7250\,K, $\log g$\,=\,1.0}\\
\hline
Gd\,{\sc ii} & 3 & 41 & 2.58 & 0.17 & 1.92\\
Yb\,{\sc ii} & 2 & 47 & 2.71 & 0.18 & 2.09\\
Lu\,{\sc ii} & 2 & ss & 1.76 & 0.03 & 2.16\\
W\,{\sc ii}  & 1 & bl & 2.92 &      & 2.27\\
\hline
\multicolumn{6}{c}{\rule[0mm]{0mm}{4mm}\irnulacht}\\
\multicolumn{6}{c}{\rule[-2mm]{0mm}{4mm}[Fe/H]\,=\,$-$0.4, T$_{\rm eff}$\,=\,7250\,K, $\log g$\,=\,1.5}\\
\hline
Gd\,{\sc ii} & 2 &  10  & 1.91 & 0.18 &   1.18 \\
Lu\,{\sc ii} & 1 &  42  & 0.97 &      &   1.30 \\
\hline
\end{tabular}
\end{center}
\vskip -.6 cm
\end{table}

\section{Discussion}
To our knowledge, this is the first systematic search for trans-Ba elements
in intrinsically enriched objects. Especially the detection of two strong Lu
lines is astonishing, but an alternative identification was not found.
\citet{tomkin83} reported the Lu\,{\sc ii} 6221.890\,\AA\ line to be blended
in the classical Ba-star HR\,774, but a blending species was, however, not
given. We think this detection is real, and that the non-detection by previous
authors is in fact a consequence of the lack of suitable line lists in the
past, the time consuming procedure of a systematic approach in this matter
and the lack of high S/N spectra obtained with efficient spectrographs on
large aperture telescopes.

\begin{figure}
\resizebox{\hsize}{!}{\includegraphics{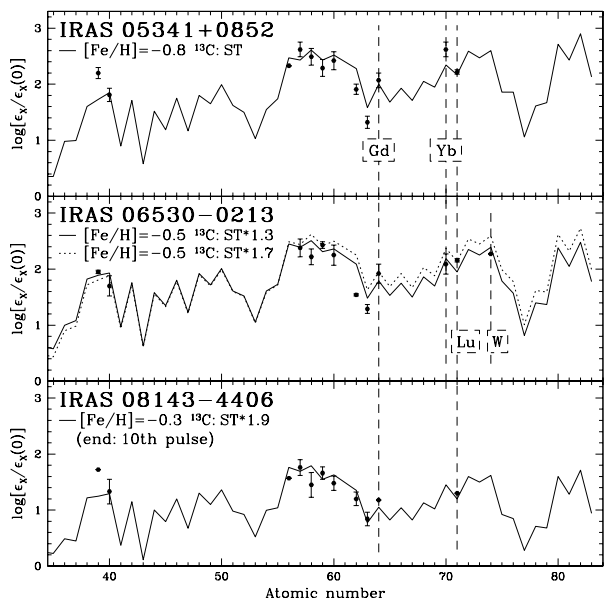}}
\vskip -.6 cm
\caption{Comparison of the observed abundances with the best fit Torino model 
predictions. The parameters of the adopted model are given in the upper left
corner. Note that the specific model is obtained by fitting the Zr and La
abundances \citep[see][ for details]{reyniers03}, so {\em without} special
attention to the very heavy s-process elements. The errorbars plotted on the
figure are the line-to-line scatters that are given in
Tab.~\ref{tab:ultrhvy050608}. The actual error on an abundance is, however,
in most cases considerably larger than this value, due to uncertainties in the
atmospheric parameters, undetected blends, uncertain solar abundances,
etc\ldots\ One example is the solar abundance of W, for which there is a
difference of 0.42\,dex between the photospheric and the meteoritic value
\citep{grevesse98}.}\label{fig:gallinoftsthlttr}
\end{figure}

As a final check for the identifications presented above, we compare the
derived abundances with the predictions of the chemical evolutionary AGB
models of the Torino group \citep[e.g.][]{gallino98, busso99, busso01}.
These models are frequently used to fit observed abundances by adjusting the
$^{13}$C pocket till the predicted and observed s-process distributions match.
The fitting procedure, together with a more general discussion of the
comparison, is described in detail in \citet{reyniers03}. Here, we focus on
the elements beyond the Ba-peak. In Fig.~\ref{fig:gallinoftsthlttr} we
graphically compare the derived abundances with the best-fit model. The
parameters of each model are given in the upper left corner; the initial mass
is 1.5\,M$_{\odot}$ for each model.

The predictions for the elements beyond the Ba-peak are consistent with the
observed abundances. This is an additional argument strengthening the line
identification for these elements. Only the Yb abundance of \irnulvijf\ is
quite off the model curve (by $+$0.3\,dex). More specifically, there is a
large difference between the Yb abundance and the Lu abundance for
\irnulvijf, which is not predicted by the models. This difference is not
seen in the abundances of \irnulzes, although three lines out of five are
in common for these elements (see Tab.~\ref{tab:ndvdllns}).

As can be seen from the comparison between two models with a different
$^{13}$C pocket for \irnulzes, the abundances of the very heavy s-process
elements are most sensitive to the adopted $^{13}$C pocket, Hence, these
elements (especially Hf and W) are ideally suited to discriminate between
possible $^{13}$C pockets. Unfortunately, a Hf abundance is very difficult
to obtain for these objects since it has only suitable lines in the blue, and
the W abundance derived from the 5104.432\,\AA\ line depends on the adopted Sm
abundance (see Sect. \ref{sect:elementsbnd}).

We conclude that we found trans-Ba abundances in strongly enriched post-AGB
stars. Despite their difficult abundance determination, they are sensitive
indicators for the strength of the neutron irradiation during the AGB
interpulse phase. Note that also in these post-AGB stars with large
overabundances of s-process elements beyond the barium stability peak, a
strong overabundance of lead (Pb, Z\,=\,82) is expected. Unfortunately Pb
has no suitable non-blended lines in the optical spectrum for these stars. 
Our detection of very heavy s-process elements indicate that also for moderate
metallicities, Pb production is expected to be significant.

\begin{acknowledgements}
Roberto Gallino is warmly thanked for providing us with his latest model
predictions, and for stimulating discussions. The anonymous referee is
thanked for the suggestion of the importance of the Lu hyperfine spitting;
Andrew McWilliam for his help with the actual hfs calculation. Both authors
acknowledge financial support from the Fund for Scientific Research - Flanders
(Belgium). This research has made use of the Vienna Atomic Line Database
(VALD), operated at Vienna, Austria, and the Database on Rare Earths At Mons
University, operated at Mons, Belgium.
\end{acknowledgements}

%\bibliographystyle{aa}
%\bibliography{Fe144.bib}
% this is the Fe144.bbl file:

\end{document}